\documentstyle[prb,preprint,aps,tighten]{revtex}
\input{psfig}
\begin{document}


\title{Diamagnetic Response of Normal-metal -- Superconductor Double Layers}

\author{W. Belzig, C. Bruder, and Gerd Sch\"on}
\address{Institut f\"ur Theoretische Festk\"orperphysik,
Universit\"at Karlsruhe, D-76128 Karlsruhe, Germany}
\date{\today}
\maketitle
\begin{abstract}
The magnetic response of a proximity-coupled superconductor-normal metal
sandwich is studied within the framework of the quasiclassical theory.
The magnetization is evaluated for finite values of the applied magnetic
field (linear and nonlinear response) at arbitrary temperatures and is used
to fit recent experimental low-temperature data. The hysteretic behavior
predicted from a Ginzburg-Landau approach and observed in experiments
is obtained within the quasiclassical theory and shown to exist also outside
the Ginzburg-Landau region.
\end{abstract}
\section{Introduction}
A superconductor in electric contact with a normal metal induces
superconducting correlations on the normal side. This
{\it proximity effect} has been studied extensively,
both theoretically and experimentally
(see Ref.~\onlinecite{deutscher:69} and references therein).
The superconducting properties of the normal metal
show up, e.g., in the conductance or the magnetization (Meissner effect).
The vanishing of the resistance of a
normal wire in contact to superconducting islands has been observed.
In recent experiments (typically with wires consisting of a
superconducting core covered by a normal metal) by
Oda and Nagano \cite{oda:80}, Mota {\it et al.}  \cite{mota:82,mota:89},
and Bergmann {\it et al.} \cite{pobell:87} the magnetization
was measured. In each of these
experiments, the magnetic field was expelled from the normal-metal part (N) of
the sample by superconducting screening currents induced by the presence of
the superconductor (S).

In the present work, we study theoretically the
proximity and Meissner effect in an NS sandwich using the
quasiclassical Eilenberger formalism \cite{eilenberger:68,larkin:68}.
In materials with a high concentration of
impurities (dirty limit), the quasiclassical theory can be
reformulated leading to the Usadel equation \cite{usadel:70}.
In both approaches the induced pair amplitude in the normal metal,
the density of states, the critical current, etc., have been calculated
(see, e.g., Refs. \onlinecite{zaikin:81,zaikin:82,kieselmann:87}).
In general, for a realistic geometry the solution of the Eilenberger or
Usadel equation together with the Maxwell equations
can be performed only numerically. We have solved the combined
system of equations in a wide range of parameters (temperature,
external magnetic field, layer thickness). In addition we
have paid special attention to the non-linear
response that shows interesting hysteretic behavior at low temperatures.

In Section II we introduce the quasiclassical Green's function formulation
for the clean and dirty limit. The geometry and the characteristic length
scales are also defined there. In Section III we evaluate the Meissner
screening current and present some results on the space dependence of the
various quantities. Finally, in  Section IV results for the susceptibility
in a wide range of temperatures and values of the magnetic field are presented
and discussed. Throughout the paper we use units with $\hbar=k_B=c=1$.

\section{The model: clean and dirty case}
Our theoretical description is based on
the quasiclassical Green's function technique \cite{eilenberger:68,larkin:68}.
We will study two limiting cases, the clean limit with complete absence of
impurities and the dirty limit with high concentration of impurities
such that the elastic scattering rate $1/\tau_{el}$ is large compared to the
superconducting order parameter $\Delta$ and the temperature $T$.
We introduce two different coherence lengths for both the
superconducting and the normal side of the double-layer structure.
In a superconducting material we have in the clean limit
\begin{eqnarray*}
	\xi^{S}_{c}
	=\frac{v_{F}}{2\Delta} \; ,
\end{eqnarray*}
and in the dirty case
\begin{eqnarray*}
	\xi^{S}_{d}
	=\sqrt{\frac{D}{2\Delta}} \; .
\end{eqnarray*}
Here $v_{F}$ is the Fermi
velocity, $\Delta$ the pair potential and $D=\frac{1}{3}v_{F}
l_{el}=\frac13v_{F}^2\tau_{el}$ the diffusion constant.  These lengths
are temperature-independent for $T\ll T_{c}$. In a normal metal we
define the temperature-dependent coherence lengths in the clean limit
\begin{eqnarray*}
	\xi^{N}_{c}(T)
	=\frac{v_{F}}{2\pi T} \; ,
\end{eqnarray*}
and in the dirty limit
\begin{eqnarray*}
	\xi^{N}_{d}(T)
	=\sqrt{\frac{D}{2\pi T}} \; .
\end{eqnarray*}

For definiteness we first consider a one-dimensional geometry with a bulk
superconductor in the region $x<0$ and a normal metal layer in the region
$0\le x\le d$ in perfect electric contact with the superconductor. (Later
we will also present some results for the cylinder geometry of the
experiments.) Since we assume $d\gg\xi^{S}_{c}$, we can
neglect the spatial dependence of the pair potential in the superconductor.
This has also been confirmed numerically \cite{higashitani:95}.
Therefore, we assume the following form for the pair potential
\begin{equation}
	\Delta(x)=\Delta\,\Theta(-x),\qquad \mbox{Im}\Delta=0\; .
\end{equation}
For the vector potential we use the Coulomb gauge. Together with the
boundary conditions it is written as
\begin{equation}
	\label{boundary.condition}
	\bbox{A}=(0,A(x),0),\quad
	A(0)=0, \quad \left.
	\frac{dA}{dx}\right|_{x=d}=H\; ,
\end{equation}
where $H$ is the external applied field. We have neglected the magnetic field
on the superconducting side, i.e., we assume that the penetration depth on the
superconducting side is much smaller than the other relevant length scales.

We first consider the clean limit defined by
\begin{eqnarray*}
	\xi^{S}_{c}, \xi^{N}_{c}(T) \ll l_{el}\; .
\end{eqnarray*}
The Eilenberger equations for this system read
\begin{equation}
	\label{eilenberger.clean}
	\left[\,(\omega+iev_yA(x))
	\hat{\tau}_{3}-
	\Delta\hat{\tau}_{1}+
	v_x\frac{\partial}{\partial x}\,
	,\,\hat{g}_\omega(v_x,v_y,x)\;\right]=0\; ,
\end{equation}
which is to be combined with the normalization condition
for the Green's functions
\begin{equation}
	\label{normalisation}
	\hat{g}_\omega^2(v_x,v_y,x)=1\; .
\end{equation}
The Matsubara frequencies are $\omega=\pi T (2n+1)$, $\hat{\tau}_i$
are the Pauli matrices and $v_x$,$v_y$ are the $x$- and $y$-components
of the Fermi velocity. We assume ideal transmission, i.e. there are no
surface potentials at the N-S-boundary and the Fermi velocities are equal
on both sides. The boundary condition reads
$\hat{g}_\omega(v_x,v_y,0-)=\hat{g}_\omega(v_x,v_y,0+)$.
At the interface to the vacuum we assume specular reflection,
$\hat{g}_\omega(v_x,v_y,d)=\hat{g}_\omega(-v_x,v_y,d)$.

The current density in the $y$-direction is given by
\begin{equation}
	\label{clean.current.general}
	j(x)=-ie2\pi N(0) T\sum_{\omega}
	<v_y\mbox{Tr}\hat{\tau}_3\hat{g}_\omega(v_x,v_y,x)>\; ,
\end{equation}
where $N(0)$ is the density of states at the Fermi level and $<\cdots>$ denotes
averaging over the Fermi surfaces that are assumed to be spherical.

In systems with high concentration of nonmagnetic impurities, such that
\begin{eqnarray*}
	\xi^{S}_{d}, \xi^{N}_{d}(T)\gg l_{el}\;,
\end{eqnarray*}
the Eilenberger equation reduces to the Usadel equation for the
isotropic part of the Green's function,
$\hat{g}_\omega(x)=\mbox{$<\hat{g}_\omega(v_x,v_y,x)>$}$. For our
geometry it can be written as
\begin{eqnarray}
	\label{usadel.equation}
	& & D\frac{d}{dx}\hat{g}_\omega(x)\frac{d}{dx}\hat{g}_\omega(x)=\\
	& & \left[\omega\hat\tau_3-\Delta(x)\hat\tau_1-De^2A(x)^2
	\hat\tau_3\hat{g}_\omega(x)\hat\tau_3\;,\;\hat{g}_\omega(x)
	\,\right] \nonumber
\end{eqnarray}
We take the same boundary conditions as in the clean case, i.~e.~we
neglect the magnetic field on the superconducting side. In the dirty
limit the current density is given by the London-like expression
\begin{eqnarray}
	\label{usadel.current}
	j(x)=\pi e^2N(0)DT\sum_{\omega}\mbox{Tr}
	\hat\tau_3\hat{g}_\omega(x)
	\left[\hat\tau_3\,,\,\hat{g}_\omega(x)\right]A(x)\,,
\end{eqnarray}
which together with the Maxwell equation
\begin{equation}
	\label{maxwell.dirty}
	\frac{d^2}{dx^2}A(x)=4\pi j(x)=\frac{1}{\lambda^2(x)}A(x)\,,
\end{equation}
defines a local penetration depth $\lambda(x)$.
\section{Meissner effect}
\subsection{Clean Limit}
In the clean limit an analytic solution of
Eq.~(\ref{eilenberger.clean}) for the
normal metal layer has been found by Zaikin \cite{zaikin:82}.
It turns out that the
$\hat{\tau}_3$ component of the Green's function in the normal metal
layer is spatially constant. Consequently the current density
(\ref{clean.current.general}) in the normal metal layer is spatially
constant as well and can be expressed as
\begin{eqnarray}
	\label{clean.jn}\nonumber
	 & & j_n=\frac{K_{c}}
	{4\pi e\xi^{N}_{c}
	(T_{c})^3}
	\frac{T}{T_{c}}\sum_{\omega>0}
	\int_0^{\frac{\pi}{2}}d\theta
	\int_0^{\frac{\pi}{2}}d\varphi\sin^2\theta\cos\varphi\sin2\phi
	\\& &
	\times\left[\left(\sqrt{1+(\frac{\omega}{\Delta})^2}
	\sinh\Phi+\frac{\omega}{\Delta}\cosh\Phi\right)^2+
	\cos^2\phi\,\right]^{-1}\; .
\end{eqnarray}
Here we have introduced a dimensionless constant
\begin{equation}
	K_c=32e^2N(0)v_{F}^2 \xi^{N}_{c} (T_{\text
	c})^2=\frac{24}{\pi} \left(\frac{\xi^{N}_{c} (T_{\text
	c})}{\lambda_{N}}\right)^2\; ,
\end{equation}
where $\lambda_{N}=(4\pi e^2N(0)n_e/m)^{-1/2}$ is the normal-metal
penetration depth, defined analogously to the London penetration depth with
the normal electron density $n_{e}$ replacing the superfluid density.
Furthermore,
\begin{equation}
	\Phi=\frac{2\omega d}{v_{F}\cos\theta}
\end{equation}
is the length of a classical trajectory divided by the thermal
coherence length $\xi^N_c(T)$ and
\begin{equation}
	\label{a-b-phase}
	\phi=2e\tan\theta\cos\varphi\int_0^dA(x)dx
\end{equation}
is the Aharonov-Bohm phase connected with this trajectory. The last
equation shows, that the  relationship between current and vector
potential is
completely nonlocal. This phase factor leads to a shift in the
energies of the Andreev levels in normal metal layer \cite{zaikin:81}.
The fact that these bound states are extended over the whole thickness
of the N-layer and that the density of Andreev levels is spatially
constant, leads to a constant current density. We can add, that
in cylindrical geometries if the normal layer is not thin compared to
the radius, this second condition is not satisfied. Hence,
as was pointed out by Nazarov \cite{nazarov:95}, the current
density is not constant in space.

To evaluate the Meissner effect, we have to solve the Maxwell
equations with the boundary condition given in
Eq.~(\ref{boundary.condition}). The solution in the N-layer is
\begin{equation}
	\label{clean.field}
	B(x)=H-4\pi j_n(d-x)\; .
\end{equation}
For the susceptibility of the N-layer we find
\begin{equation}
	\label{susc.clean}
	\chi=-\frac{j_n d}{H}\; .
\end{equation}
Since the solution of the problem may lead to $4\pi j_n d>H$
the magnetic field in the region $0\le x\le d-H/4\pi\j_n$ may change its
sign relative to the applied field. This
overscreening effect was first found by Zaikin \cite{zaikin:82}.
In Fig.~\ref{clean.current} we have plotted the current density and
the magnetic field in the N-layer for different temperatures in the
limit of small magnetic fields. Below $T\sim 0.1 T_c$
we find the overscreening. Below this
temperature the screening quickly reaches its saturation value
(solid curve) and a susceptibility of $3/4$ as compared to
complete screening. Thus, even at $T=0$ the screening is incomplete.

Next we study the nonlinear response. For this purpose we write the
self-consistency equation for the integrated dimensionless vector potential
$a=e\int_{0}^{d}A(x)dx$
\begin{equation}
	\label{vp.integral}
	a=e\frac12Hd^2-\frac{4\pi}{3}d^3ej_n(a)\;.
\end{equation}
Solving Eq.~(\ref{vp.integral}) for the applied field $H$ we get,
together with Eq.~(\ref{susc.clean}), the magnetization curve
$\chi(H)$
\begin{eqnarray}
	\chi(a)&=&-\frac{j_{n}(a)\,d}{2 H(a)}\\\nonumber
	H(a)&=&\frac{2a}{ed^2}+4\pi j_{n}(a)\frac{2d}{3}
\end{eqnarray}
parameterized by the integrated vector potential.
\subsection{Dirty Limit}
Here Eq.~(\ref{usadel.equation}) cannot be solved analytically anymore.
To proceed numerically we reduce it to the scalar equation
\begin{eqnarray}
	\label{usadel.scalar}
	& & 2\omega F_\omega(x)-2\Delta G_\omega
	-4 D e^2A^2(x)F_\omega(x)G_\omega(x)= \nonumber \\ & &
	D\left[G_\omega(x)\frac{d^2}{dx^2}F_\omega(x)-
	F_\omega(x)\frac{d^2}{dx^2}G_\omega(x)\right],
\end{eqnarray}
where $F_\omega(x)=\frac12\mbox{Tr}\hat\tau_1\hat{g}_\omega(x)$ is the
off-diagonal part of the Green's function and
$G_\omega(x)=(1-F_\omega(x)^2)^{1/2}$ is fixed by the normalization
(\ref{normalisation}).  The Maxwell equation is given by
Eq.~(\ref{maxwell.dirty}) with the boundary conditions given in
(\ref{boundary.condition}).  We write for the local penetration depth
\begin{equation}
	\label{local.pendepth}
	\frac{1}{\lambda^2(x)}=
	\frac{K_d}{\xi^{N}_{d}(T_{c})^2}
	\frac{T}{T_{c}}\sum_{\omega>0}F_\omega^2(x)\;,
\end{equation}
where we have defined an dimensionless constant $K_{d}=12(l_{\text
el}/\pi \lambda_{N})^2$.

We have solved this system of equations
numerically for a range of parameters.
Some results are shown in Fig.~\ref{dirty.current} for a layer
thickness of $50\xi^{N}_{d}(T_c)$ and $K_{d}=100$ in a weak magnetic
field, where we neglect the term $\sim A^2$ in the Usadel equation. At
the highest temperature, $T=0.3T_{c}$, the field penetrates through
the whole N-layer. Near the N-S-boundary, where the induced
superconductivity is strong, the field decays rapidly to zero. At lower
temperatures the field-free region increases, because the induced
superconductivity extends to a region of linear size $\sim
\xi^N_d(T)$. At the lowest temperature shown here, $T=0.001 T_{c}$,
the field expulsion is London like and the field as well as the
current density fall off exponentially. The reason is that at
this temperature the local penetration depth near the
N-Vacuum-boundary is practically constant. In the intermediate
temperature regime the current flows in a well defined region inside
the N-layer, and consequently the screening takes place in this
region.

To calculate the nonlinear response we have to solve the complete
Usadel equation Eq.~(\ref{usadel.equation}) together with the Maxwell
equations including the local penetration depth (\ref{local.pendepth})
numerically. From the solution of the Maxwell equations, we
find the susceptibility of the normal-metal part
\begin{equation}
	\chi=-\frac{1}{4\pi}\left(\frac{A(d)}{Hd}-1\right)\; .
\end{equation}
Results for the clean and the dirty case are discussed in the next
section.
\section{Susceptibility}
\subsection{Linear Response}
First we study the case of linear response. In the clean limit
we linearize Eq.~(\ref{clean.jn}) with respect to the Aharonov-Bohm
phase $\phi$. The resulting expression has been analyzed analytically
by Zaikin in some limits \cite{zaikin:82} and numerically by
Higashitani and Nagai \cite{higashitani:95}. We have performed the
remaining integration over $\theta$ numerically for all temperatures.
In Fig.~\ref{clean.linear} some results for different layer
thicknesses are shown. The temperature at which screening sets in, as
well as the slope of the temperature dependence, changes drastically
when the layer thickness increases. At $T=0$ the integrals can be
solved analytically. For $d\gg\lambda_{N}$, the saturation value is
equal to $3/4$ of a perfect diamagnet, independent of the thickness
\cite{zaikin:82}.

Here we should mention that the temperature dependence of the
susceptibility cannot be fitted to a power law $\chi \propto
T^{-\alpha}$ in any temperature range, in contrast to earlier
theoretical predictions \cite{orsay:66} which suggest powers in the
range $1/2\le\alpha\le 1$ depending on the impurity
concentration. Rather it satisfies an exponential law. This explains
qualitatively some experimental results \cite{mota:82,mota:89,pobell:87},
where exponents $\alpha$ with values up to $2$ had been found.

In the dirty limit, we neglect the term $\sim A^2$ in
Eq.~(\ref{usadel.equation}) and solve the resulting differential
equation numerically.  In Fig.~\ref{dirty.linear} some results of our
calculation are shown. In comparison with the clean case, the
screening sets in at higher temperatures and the saturation values can
be larger (depending on $K_{d}$). The temperature dependence is well
described by
\begin{equation}
	\chi\sim (T^{-1/2}-const)
\end{equation}
in the intermediate temperature regime well below $T_{c}$ and above
the saturation temperature.  Our results agree with previous works in
the applicable limits. In earlier work \cite{orsay:66}, a generalized
Ginzburg-Landau approach was used.  Narikiyo and Fukuyama
\cite{narikiyo:89} linearized Eq.~(\ref{usadel.equation}) with respect
to $F$ and solved the system of equations for an infinite system
numerically.  However, our calculation is free of the limitations of
the Ginzburg-Landau theory and is valid at any temperature.
Furthermore, we have taken into account the nonlinearity of the Usadel
equation and finite-size effects.

The behavior shown in Fig.~\ref{dirty.linear} has also been observed
experimentally in very dirty samples \cite{oda:80,pobell:87}.
In these experiments, a temperature dependence described by
$\chi\sim(T^{-1/2}-const)$ was found and saturation values of the
susceptibility of $90-95\%$ of a perfect diamagnet.

In Fig.~\ref{dirty.exp} we show a comparison between theory and
experiment. The experimental data \cite{mota:data} are measured on a
cylindrical geometry with a superconducting core surrounded by a
normal metal layer of thickness $d$. The theory has been generalized
to this kind of geometry and also been solved numerically. The
parameters for the theory were calculated from experimental values for
$d$ and $l_{el}$ and from theoretical values for $v_F$ and $\lambda_N$. No
fitting parameter was used. The agreement is quite remarkable at low
temperatures. At higher temperatures ($T\stackrel{>}{\sim}0.1T_c$) the
dirty limit theory cannot be applied anymore and this explains the
disagreement in this temperature regime.
\subsection{Nonlinear Response}
We will now turn to the nonlinear response. In
Fig.~\ref{clean.nonlinear} some numerical results for the clean case
are shown. At the highest temperature shown here, the susceptibility
drops to zero around $h\approx 0.02$. At lower temperatures
the solution is not unique anymore.  There is one solution with the
maximal field expulsion, the value of which depends on temperature
(lower thick lines), and one solution, for which the field completely
penetrates in the N-layer and there is no screening (upper thick
lines).  The third solution with the negative derivative is unstable
(thin lines).
The region in which the solution is non-unique grows,
as the temperature is lowered, but the lower boundary of this region
stays at the same value of the field. Since only one solution can be
stable, there must be a jump in the susceptibility at a certain
point. In principle we could determine this point by comparing
 the free energies of the N-layer for the two different solutions.
Unfortunately the free-energy
functional proposed by Eilenberger \cite{eilenberger:68} cannot be
applied for $\Delta=0$ (normal region).

In experiments the situation is often quite different.
On lowering or raising the field, the jump in the
susceptibility would not occur at the value predicted theoretically,
but effects like ``superheating'' or ``supercooling'' are likely to
occur. These effects were observed in very clean samples
\cite{mota:89}. When the field is increased, the susceptibility will
remain on the lower branch up to a certain field value and then will
jump to the upper branch. On the other hand, when the field is
decreased, the system will stay on the upper branch with complete
field penetration. At a certain value of the field, the field will
suddenly be expelled and the system will jump to the lower branch. These
jump fields do not necessarily coincide with the boundaries of the
instability zone.

The boundaries of the hysteretic regions are shown in
Fig.~\ref{clean.breakdown} for different layer thicknesses. The
superheating field depends exponentially on temperature, whereas the
supercooling field depends not very strongly on temperature. At
low temperatures the superheating field tends to saturate at a
certain value, which is independent of the layer thickness. Further
calculations show that the saturation value is proportional to the material
constant $K_{c}$.

In the dirty limit there is also a temperature-dependent hysteretic
effect, as shown in Fig.~\ref{dirty.nonlinear}. Above temperatures of
about $0.03 T_{c}$ the magnetization curve is unique. Below these
temperatures there is a region with constant lower boundary and
increasing upper boundary as in the clean case. The absolute values of
the susceptibility are drastically different. The saturation value at
low fields and low temperatures can reach complete field expulsion for
appropriately chosen values of $K_d$.  On the other hand the jump in
the susceptibility is less than in the clean case. On increasing the
field the susceptibility is reduced only by $\approx 50\%$ at the jump
and reaches slowly zero, if the field is increased further. On
lowering the field, the jump is a slightly smaller. This hysteretic
behavior was observed \cite{pobell:87}.

\begin{figure}
	\caption[]{\label{dirty.breakdown} Limits of the instability
	region (dirty case): supercooling field $h_{sc}$ (open
	symbols) and superheating field $h_{sh}$ (filled symbols) for
	$K_{d}=10$.  $h$ is defined as in the previous figure.}
\end{figure}
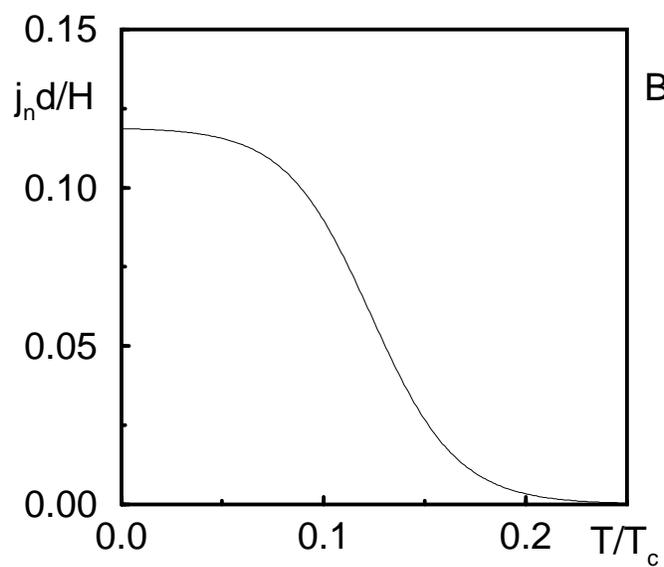
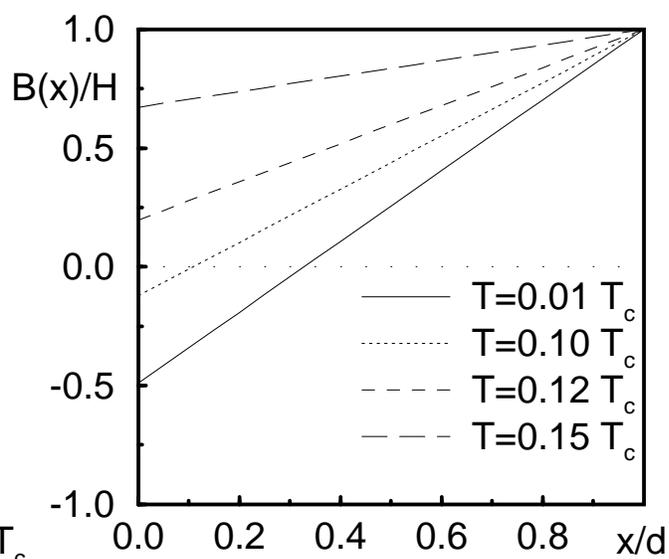
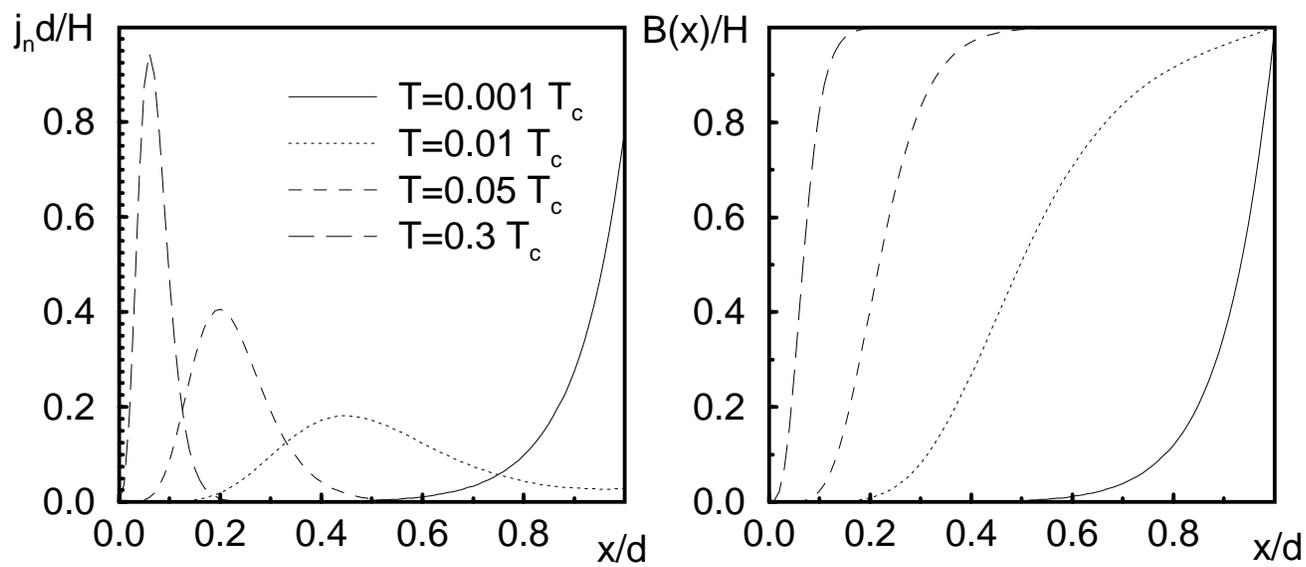
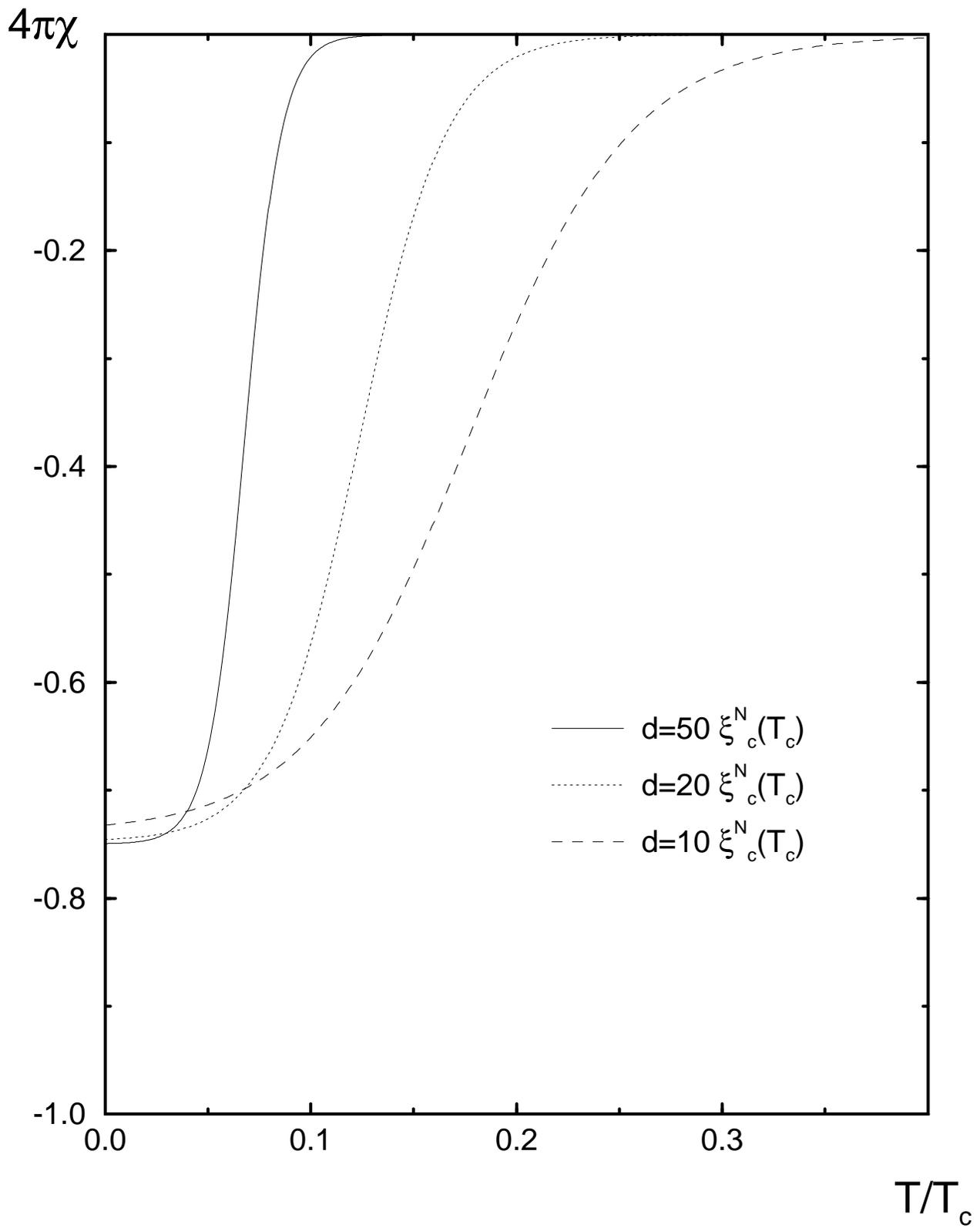
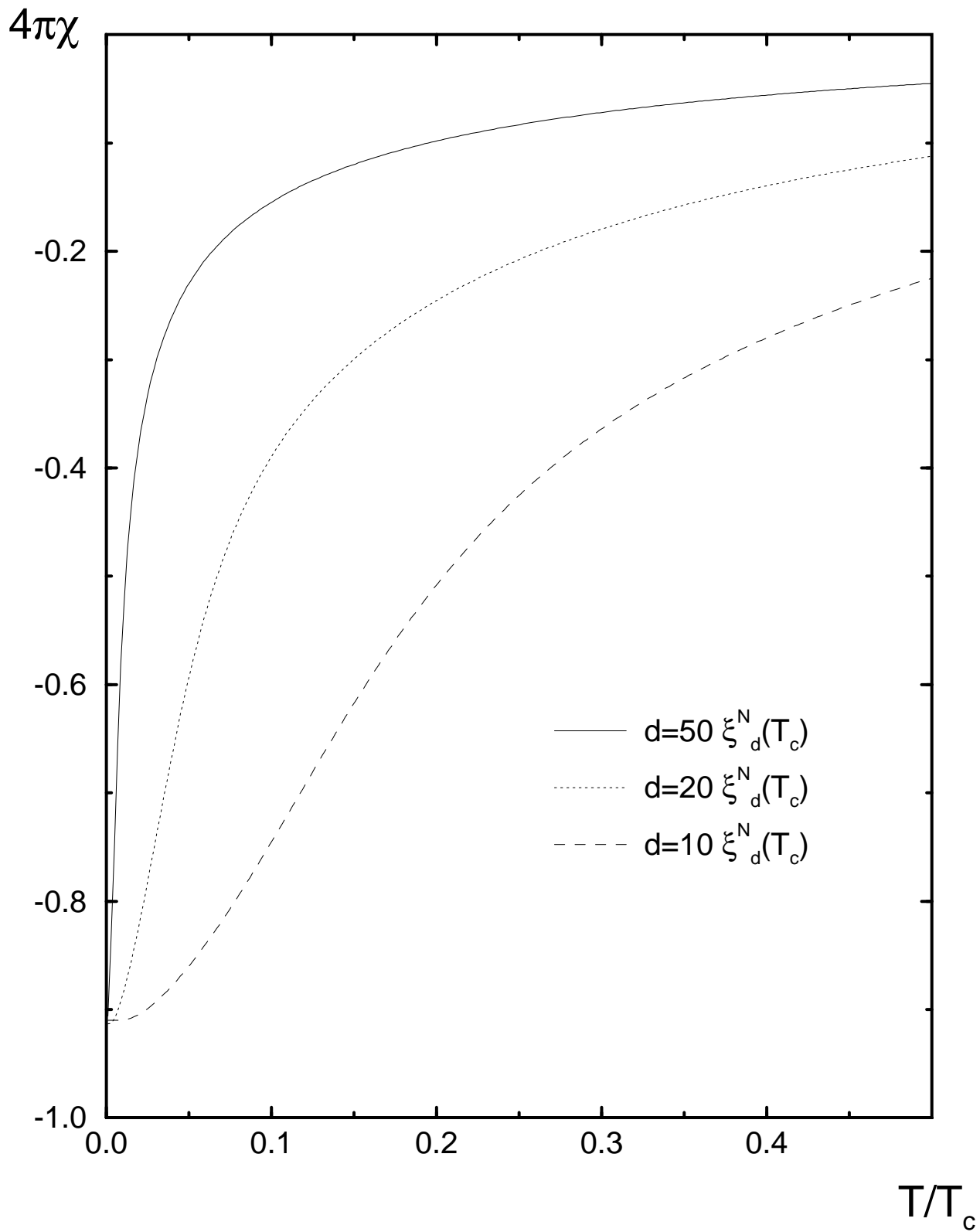
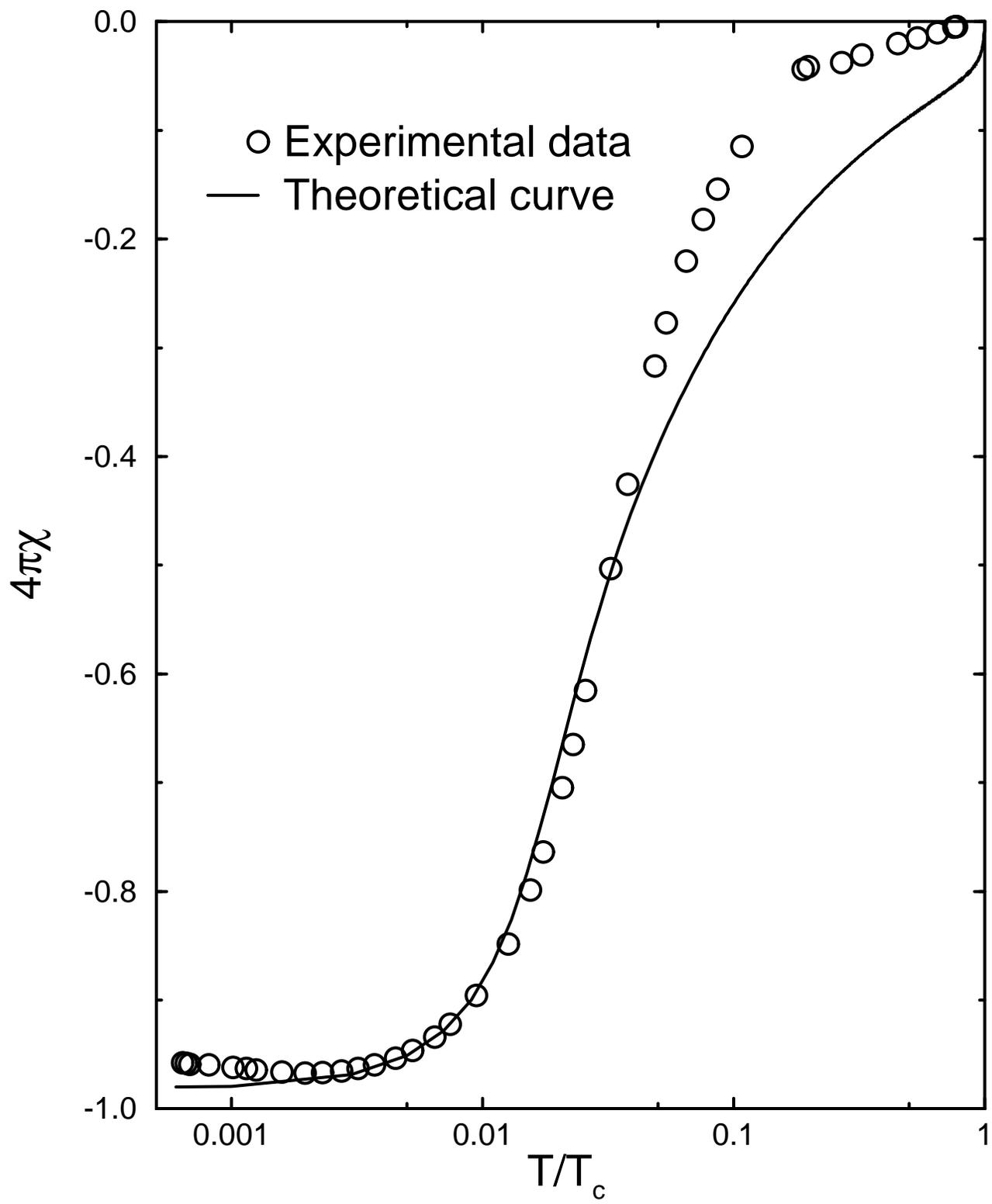
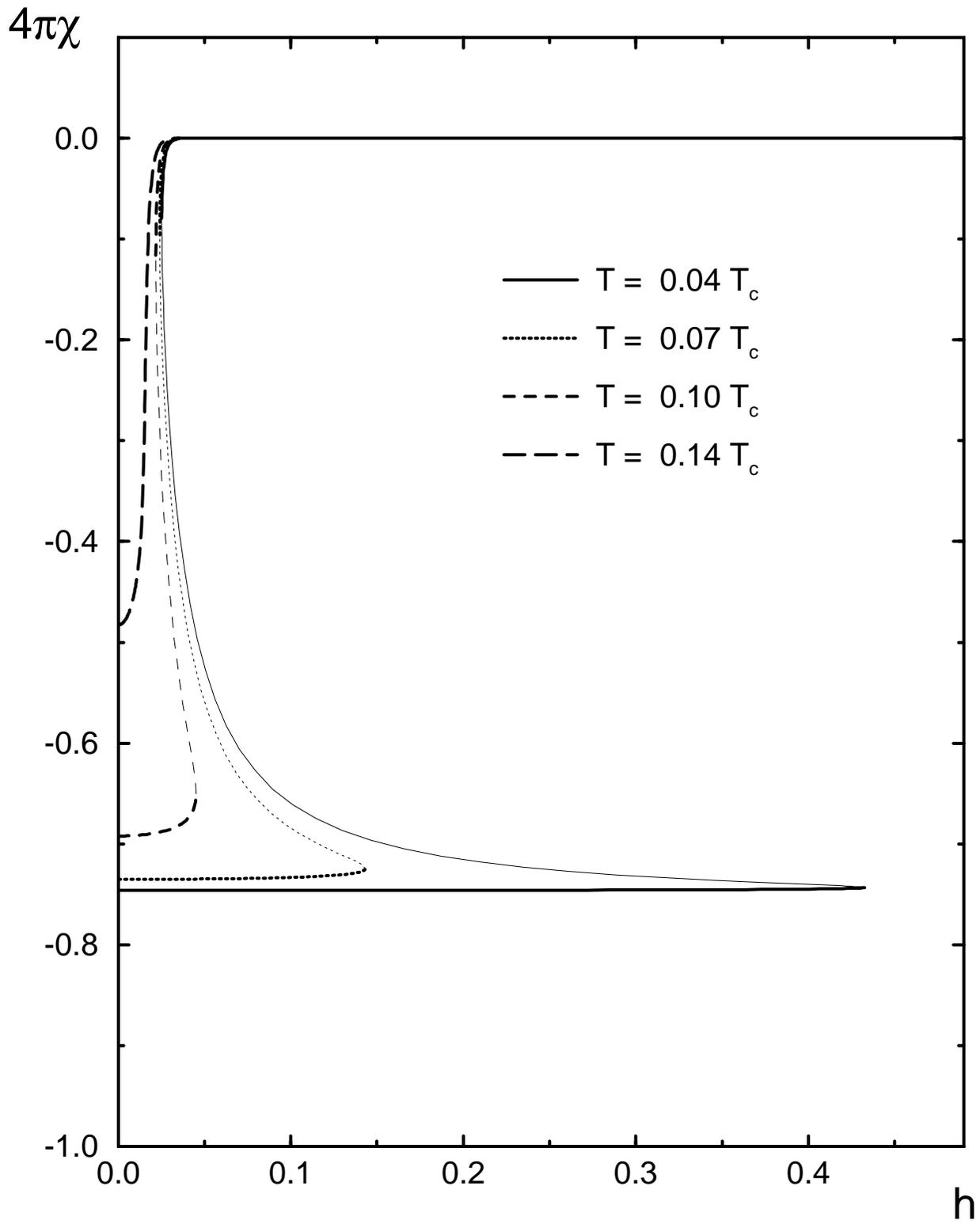
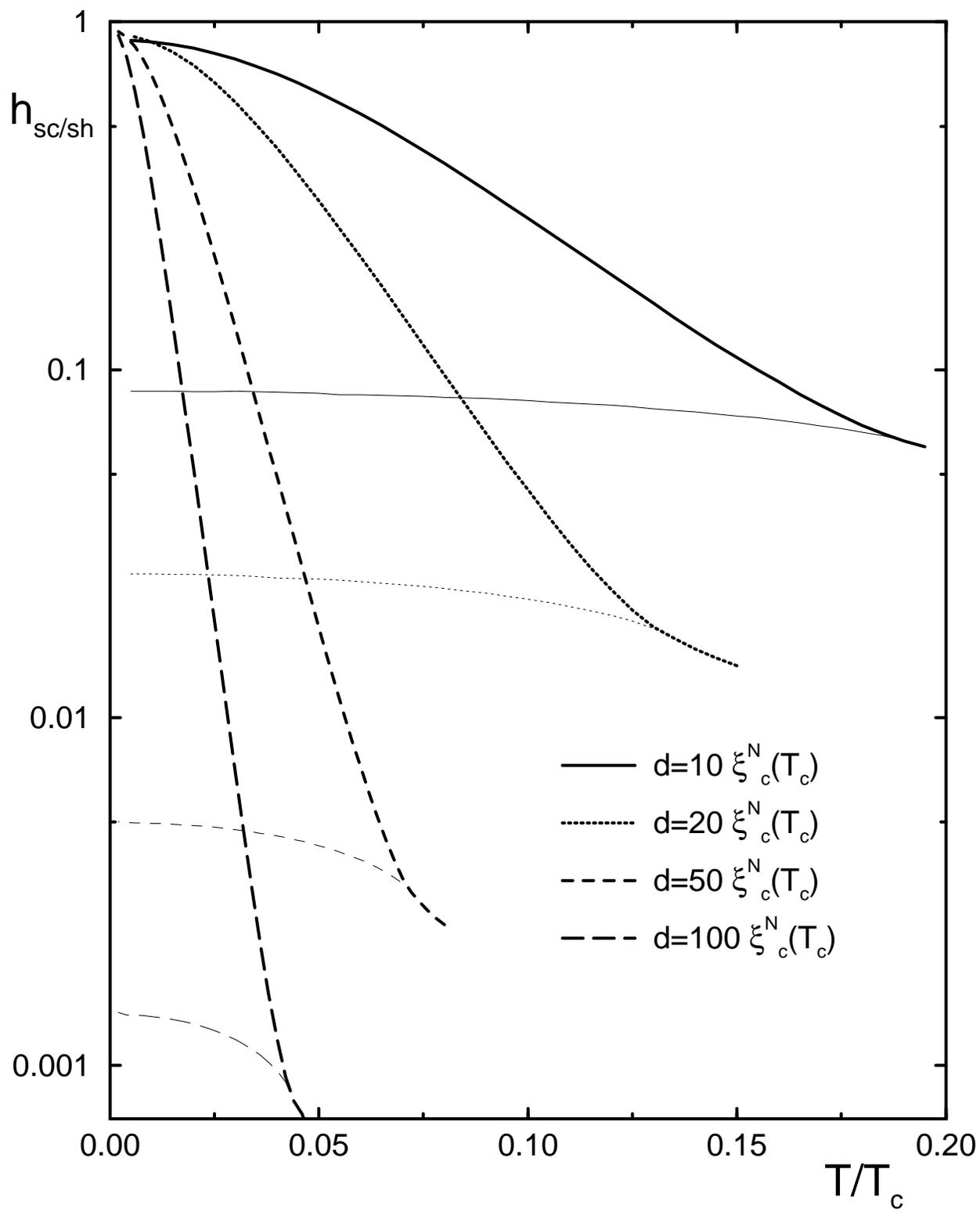
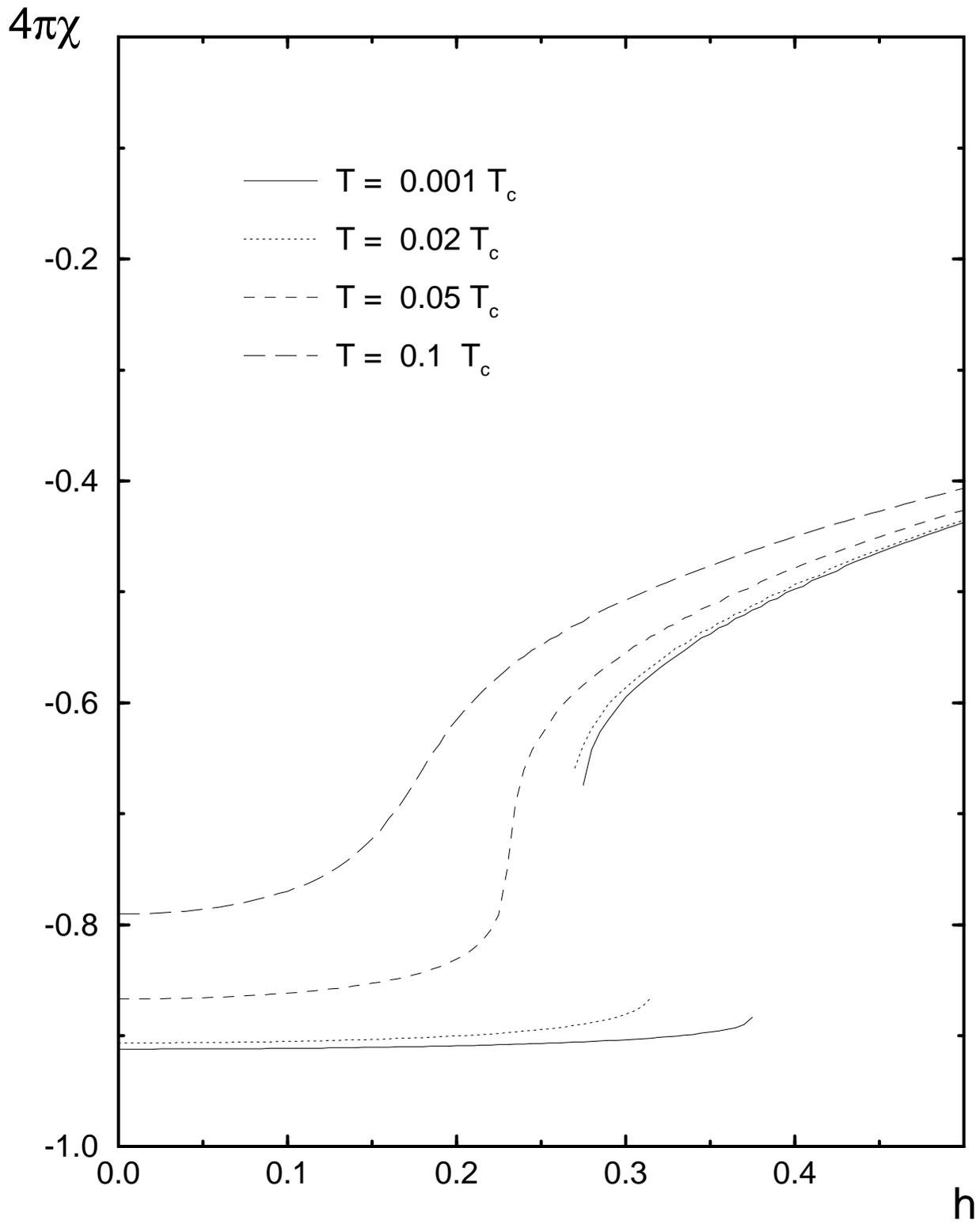
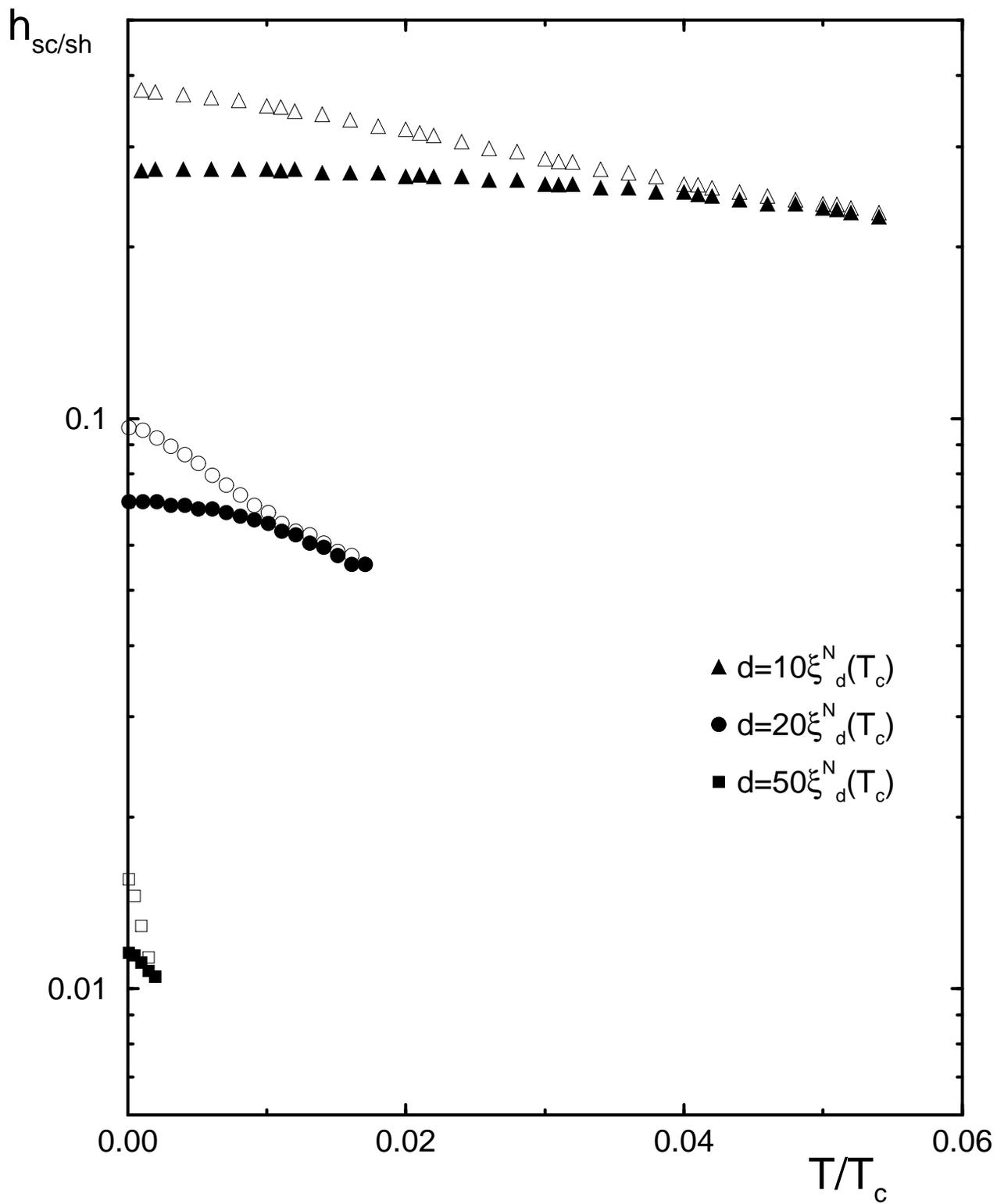

In Fig.~\ref{dirty.breakdown} the limits of the region with hysteretic
behavior for different layer thicknesses in the dirty limit are
shown. As in the clean case the superheating field depends
exponentially on the temperature, but not as strongly.  Also, the
supercooling field tends to become a constant at low temperatures and
does not very strongly depend on temperature.  In contrast to the
clean case the saturation values of both fields at low temperatures
depend strongly on the layer thickness.

Finally, we would like to comment on some of our assumptions. We have
assumed ideal interfaces between normal metal and superconductor,
specular reflection at the interface between normal metal and vacuum,
and spherical Fermi surfaces in both materials. The formalism
presented in this paper can be modified to describe more general
physical situations. Including
e.g. non-ideal interfaces will weaken the proximity effect and
diminish the diamagnetic response of the normal metal. In this paper,
we have concentrated on presenting a model calculation in which the
main aspects of the diamagnetic response of NS sandwiches can be
studied. Despite our simplifying assumptions, the calculation was
shown to be in reasonable agreement with experiment, see Fig.~\ref{dirty.exp}.
Further calculations will be necessary to describe the high-temperature
behaviour correctly and to study arbitrary concentrations of impurities,
i.e., cases that are neither clean nor dirty.

In conclusion, we have applied the quasiclassical (Eilenberger or
Usadel) theory to calculate the Meissner effect in a proximity
sandwich. We have evaluated the full non-linear magnetic response at
all temperatures and for various thicknesses of the normal layer. We
have shown that both in the clean and in the dirty limit a hysteretic
behavior of the magnetization of a normal metal layer in proximity
with a superconductor is possible, as has been observed experimentally
in both limits.

\acknowledgements We would like to thank A.~C. Mota, Yu.~V. Nazarov,
and A.~D. Zaikin for helpful discussions, and R. Frassanito for
providing the data shown in Fig. \ref{dirty.exp}.

\begin{figure}
	\caption[]{\label{clean.current} Current density and magnetic
	field in the N-layer (clean limit). Layer thickness $d=20\xi^{\text
	N}_{c}(T_c)$, dimensionless material constant $K_{c}=10$.}
\end{figure}

\begin{figure}
	\caption[]{\label{dirty.current} Current density and magnetic
	field in the N-layer (dirty limit). Layer thickness
	$d=50\xi^{\text N}_{d}(T_{c})$ and dimensionless material
	constant $K_{d}=100$.}
\end{figure}
\begin{figure}
	\caption[]{\label{clean.linear} Linear susceptibility (clean case)
	for $K_{c}=10$.}
\end{figure}
\begin{figure}
	\caption[]{\label{dirty.linear} Linear susceptibility
	(dirty limit) for $K_{d}=100$.}
\end{figure}
\begin{figure}
	\caption[]{\label{dirty.exp}Comparison between experiment
	(AgNb)\protect\cite{mota:data} and dirty limit theory. Experimental
	parameters: Radius of the S-core$=35\mu$m, $d=14.5\mu$m, and
	$l_{el}=2.0\mu$m. In the theory we have used
	$d=41.5\xi^N_d(T_c)$ and $K_d=10000$.}
\end{figure}
\begin{figure}
	\caption[]{\label{clean.nonlinear} Nonlinear susceptibility
	(clean case) of an N-layer of thickness
	$20\xi^{N}_{c}
	(T_{c})$ and $K_{c}=10$
	as a function of the applied field for various
	temperatures. The dimensionless field is defined by $h=H
	\pi\xi^{N}_{c} (T_{c})^2/\Phi_0$. }
\end{figure}
\begin{figure}
	\caption[]{\label{clean.breakdown} Limits of the instability
	region: supercooling field $h_{sc}$ (thin lines) and
	superheating field $h_{sh}$ (thick lines) for $K_{c}=10$.
	$h$ is defined as in the previous figure.}
\end{figure}
\begin{figure}
	\caption[]{\label{dirty.nonlinear} Nonlinear susceptibility
	(dirty case) of an N-layer of thickness
	$10\xi^{N}_{d}
	(T_{c})$ and $K_{d}=100$
	as a function of the applied field for various
	temperatures. The dimensionless field is defined by $h=H
	\pi\xi^{N}_{d} (T_{c})^2/\Phi_0$. }
\end{figure}
\end{document}